# A high throughput Intrusion Detection System (IDS) to enhance the security of data transmission among research centers

**M. Grossi,**[a,b,1] **F. Alfonsi,**[a,c] **M. Prandini,**[d] **and A. Gabrielli**[a,c]

[a] *Department of Physics and Astronomy "Augusto Righi" (DIFA), Viale Carlo Berti Pichat, 6/2, 40127 Bologna, Italy*

[b] *INFN-CNAF Bologna, Viale Carlo Berti Pichat, 6, 40127 Bologna, Italy*

[c] *INFN Bologna, Viale Carlo Berti Pichat, 6/2, 40127 Bologna, Italy*

[d] *Department of Computer Science and Engineering Viale del Risorgimento, 2, 40136 Bologna, Italy*

  *E-mail*: `marco.grossi8@unibo.it`

ABSTRACT: Data breaches and cyberattacks represent a severe problem in higher education institutions and universities that can result in illegal access to sensitive information and data loss. To enhance the security of data transmission, Intrusion Prevention Systems (IPS, i.e., firewalls) and Intrusion Detection Systems (IDS, i.e., packet sniffers) are used to detect potential threats in the exchanged data. IPSs and IDSs are usually designed as software programs running on a server machine. However, when the speed of exchanged data is too high, this solution can become unreliable. In this case, IPSs and IDSs designed on a real hardware platform, such as ASICs and FPGAs, represent a more reliable solution. This paper presents a packet sniffer that was designed using a commercial FPGA development board. The system can support a data throughput of 10 Gbit/s with preliminary results showing that the speed of data transmission can be reliably extended to 100 Gbit/s. The designed system is highly configurable by the user and can enhance the data protection of information transmitted using the Ethernet protocol. It is particularly suited for the security of universities and research centers, where point-to-point network connections are dominant and large amount of sensitive data are shared among different hosts.

KEYWORDS: Network security; packet sniffer; data protection, FPGA, embedded systems.

## Contents

## 1. Introduction

With the advent of the Internet of Things (IoT) and the interconnection of embedded technologies, gigabits of personal data (including, names, addresses, driver's license numbers, social security numbers, etc.) are stored by different institutions such as governments, universities, and hospitals [1]. The increasing complexity of networks due to the increase of web applications and business transactions, results in high vulnerability to data breaches that can result in denial of service, unauthorized accesses, and sensitive data theft or alteration. This causes severe concerns to organizations that operate using network technology and must spend time finding ways to protect themselves from data breaches [2].

    Cybersecurity in science plays a crucial role in protecting sensitive data, research findings, intellectual property, and infrastructure within the scientific community. As technology advances and more scientific processes and data are digitized, the need for robust cybersecurity measures becomes increasingly important. Cyberattacks and data breaches have been reported in higher education institutions and universities [3]. For example, in 2014 the University of Maryland network was hacked and the personal data of over 300,000 students were illegally accessed, with a cost for the university of over 6 million USD. In the same year, personal data (including social security numbers and academic records) of over 30,000 students were violated at the Riverside Community College District in California.

    Protecting scientific data with a firewall is an essential aspect of ensuring data security and preventing unauthorized access to sensitive information [4]. Intrusion Prevention Systems (IPSs, i.e., firewalls) and Intrusion Detection Systems (IDSs, i.e., packet sniffers) act as a barrier between a trusted internal network and untrusted external networks, such as the Internet. These systems can be used in different places within a network and the choice of their placement depends on the network architecture, security requirements, and the level of control needed for the network traffic [5].

IPSs and IDSs are often implemented in software running on a server machine [6,7,8]. However, when the amount of transferred data and the speed of data transfer increase over a certain threshold, these systems can lose effectiveness. In these situations, a hardware implementation is preferred since this can guarantee real-time operations and much higher data throughput [9, 10].

Network security systems can work according to two different principles:

1) IPS, where network data are filtered according to a set of rules, allowing to pass only the data that meet the rule's requirements. In this case, the firewall features two different ports, working in full duplex and when network data are presented at the input of one port, these are transferred to the output of the other port if no threats are detected or blocked, otherwise.

2) IDS, working as a passive device that reads all the traffic and sends alarm messages to a remote server when potential threats are detected [11, 12].

In this paper, a packet sniffer was designed in Verilog with the Vivado IDE, using a FPGA development board, and its performance was tested using an ad-hoc designed packet generator deployed using another FPGA development board. The designed packet sniffer can support a data rate of up to 10 Gbit/s with preliminary results that extend its performance to 100 Gbit/s. It analyses Ethernet packets of type ARP, IP, TCP, UDP, and ICMP and can generate an alert message when packets that do not match a set of rules defined by the user are read. The designed device is dynamic, in that it can be programmed at all firmware levels and can constitute not only an information security element for an intelligent network system, but also an element of symbiotic exchange of skills between universities, industry, and research centers. In fact, for research institutions such as CNAF-CERN that need huge data exchange between remote machines (mirrors), the possibility of increasing security from intrusions and attacks also coming from the Internet is today of increasingly fundamental importance. In Section 2 the proposed packet sniffer and the packet generator used to test its performance are discussed in detail. In Section 3 the results of the performance test of the proposed 10 Gbit/s packet sniffer are presented and preliminary results to extend the proposed packet sniffer for higher data rates (100 Gbit/s) are discussed. Finally, in Section 4 concluding remarks are presented.

## 2. Experimental setup

### 2.1 Packet generator

The packet generator is implemented using a KC705 FPGA development board by Xilinx [13]. This development board integrates a Kintex7 XC7K325T-2FFG900C FPGA device (326080 logic cells, 840 DSP slices, 16 Gigabit transceivers, 500 I/O pins) and features, among other things, different clock sources, 1GB DDR3 RAM, 16MB SPI Flash memory, RJ45 and small form-factor pluggable (SFP) connectors for network interfacing, 8 lane PCI Express interface, 2x16 LCD display and USB UART for data communication with a PC.

The packet generator is implemented in the programmable logic of the FPGA device. It can generate Ethernet packets of type ARP, UDP, TCP, and ICMP with a maximum data rate of 10 Gbit/s. The packets to be generated are transferred from a PC running LabVIEW (National Instruments) programs to the FPGA memory using UART communication at a 115200 baud rate. Once the FPGA memory is loaded with the packet data, the user can set different commands using the UART, such as to generate a single loop of packets, to generate the packets continuously, and to stop the packet generation. During the packet generator programming, the user can also set the number of clock cycles of delay between one packet and the next one.

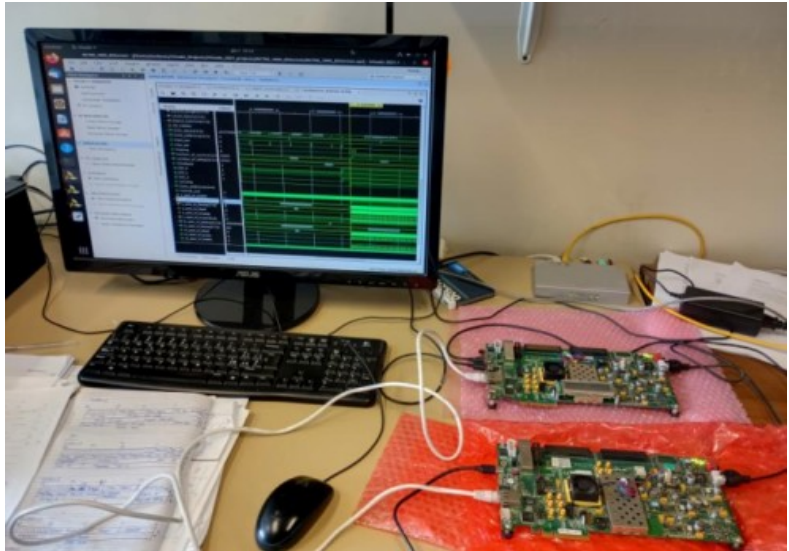

**Figure 1.** Experimental setup used to validate the proposed packet sniffer and test the system performance.

The whole system composed of the packet generator, the packet sniffer and the PC used for FPGA programming and data transfer using UART is presented in Fig. 1.

The packet generator implemented in the FPGA programmable logic uses three different clock sources: a 200 MHz clock source to set the properties of the AXI bus interface, a 156.25 MHz clock that provides the timing for the Ethernet packet generation and a 10 MHz clock that provides the timing for the UART communication. Once the packet data are fetched from the FPGA block memory, data are provided as input to the 10G Ethernet Subsystem, an IP module provided by Xilinx that manages the physical and data link layers of the Ethernet protocol, and Ethernet data are generated in serial differential format at the SFP connector present on the KC705 board. At the input of the 10G Ethernet Subsystem, data are handled with 8-byte (64 bits) parallelism and synchronized with a 156.25 MHz clock (period of 6.4 ns). The signal waveforms resulting from a Vivado simulation in the case of the generation of a single UDP packet of length 74 bytes are presented in Fig. 2.

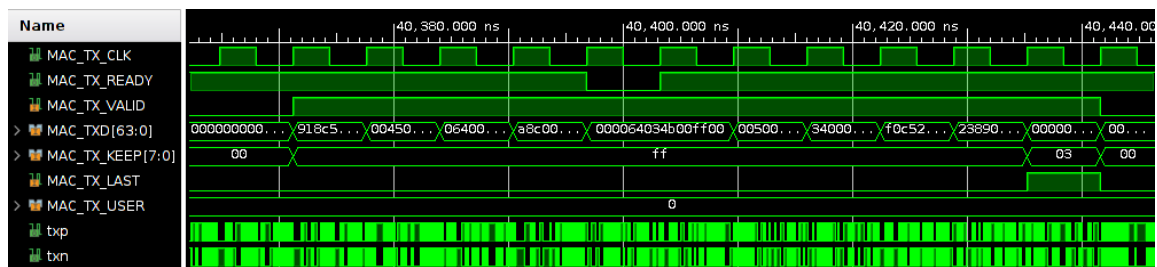

**Figure 2.** Signal waveforms from a Vivado simulation in the case of generation of a single UDP packet of length 74 bytes.

In Fig. 2 MAC_TX_CLK is the 156.25MHz clock, MAC_TX_VALID is a control signal that is active only during data transmission, MAC_TX_READY is a signal that is enabled when the system is ready to sample the data, MAC_TXD is the 64 bit segment of the packet data, MAC_TX_KEEP is a 8 bit signal used to define how many bytes of MAC_TXD contain valid data, MAC_TX_LAST is a signal enabled only during the transmission of the last 64 bit word and MAC_TX_USER is a signal used to define the presence of errors. Since the minimum length of an Ethernet frame is 64 bytes, the transmission of a single Ethernet frame needs a minimum of 8 clock cycles.

## 2.2 Packet sniffer

The packet sniffer has been designed using a second KC705 development board. The schematic of the system is represented in Fig. 3.

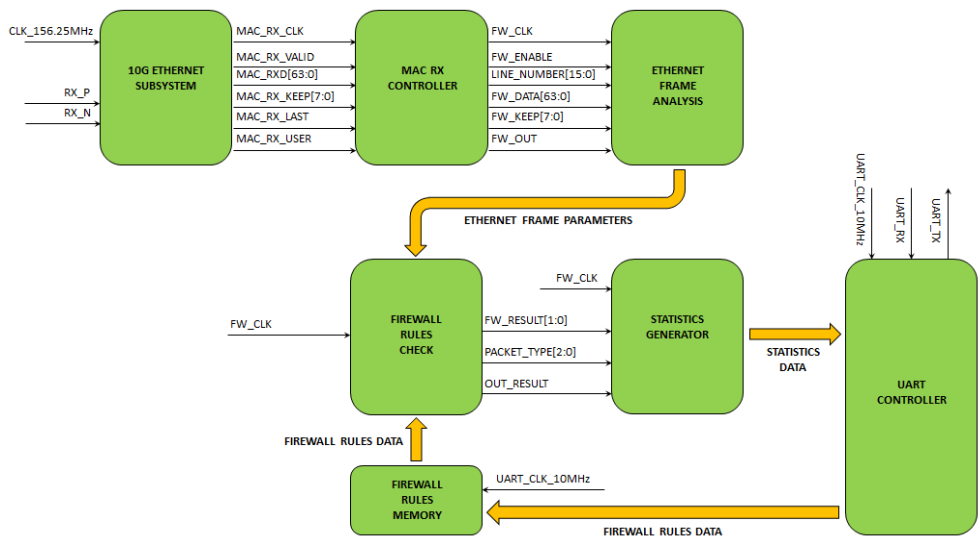

**Figure 3.** Schematic of the hardware modules that compose the designed packet sniffer.

Ethernet data are fed to the system using the differential signals RX_P and RX_N (with the SFP connector integrated into the KC705 board) and are transformed to parallel data (64-bit width) by the 10G Ethernet Subsystem IP module. Data are then fed to the MAC RX CONTROLLER module that generates all the control and data signals for the ETHERNET FRAME ANALYSIS module where the header parameters of the Ethernet frame (such as MAC source and destination addresses, IP source and destination addresses, source port, and destination port) are calculated. The header parameters are used by the FIREWALL RULES CHECK module to evaluate if the Ethernet frame represents a potential threat to the network, based on a set of rules that are stored in the FIREWALL RULES MEMORY module, composed of four block RAM modules each storing four words of 224 bit, and loaded by the user through the UART interface. The packet sniffer rules are based on a whitelist approach and check the Ethernet frame header parameters against allowed values of the IP source and destination addresses, source and destination ports, and data protocols. The result of the rules check is sent to the STATISTICS GENERATOR module where a set of statistics (such as the number of allowed packets, and number of packets of protocol UDP, TCP, ICMP) are calculated and transferred to the PC using the UART interface for further processing and data filing.

The signal waveforms resulting from a Vivado simulation in the case of the Ethernet frame header parameters calculation are presented in Fig. 4.

The waveforms presented in Fig. 4 are for the case of four Ethernet frames of level 3 protocol IP (0x0800), two of level 4 protocol UDP (17), and two of level 4 protocol TCP (6).

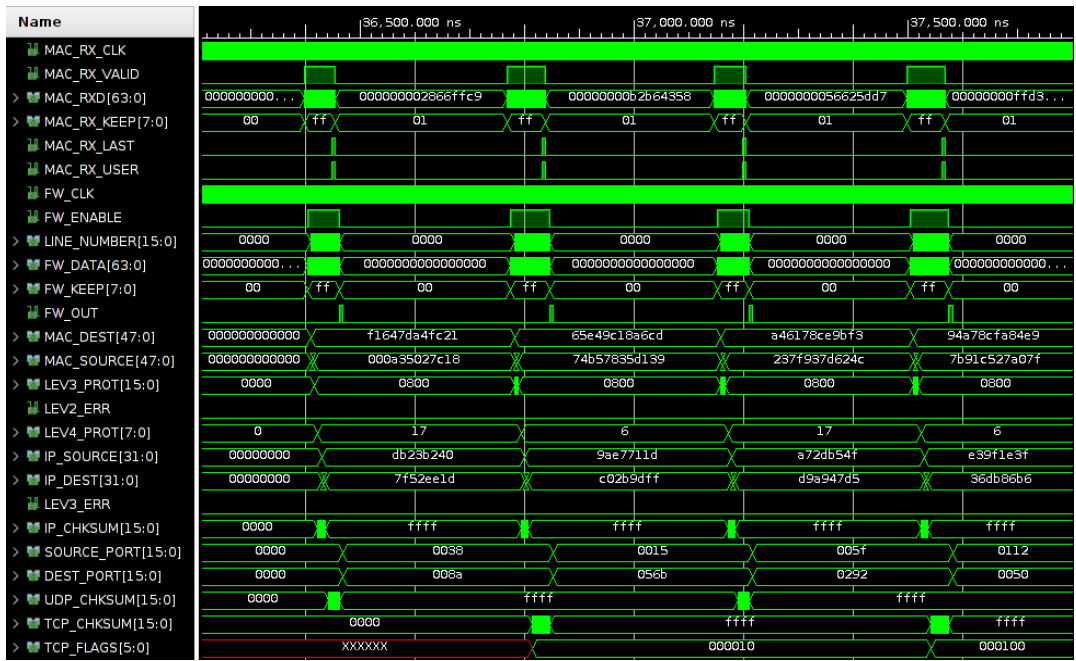
**Figure 4.** Signals waveform in the case of Ethernet frame header parameters calculation.

For each Ethernet frame the checksums for level 3 and level 4 are calculated to detect the presence of data corruption. The calculated header parameters are valid and can be sampled by the next module when the signal FW_OUT is enabled. The signal waveforms for the evaluation stage of the Ethernet frame compliance to the set of packet sniffer rules and the statistics calculation are presented in Fig. 5.

As can be seen, the signal FW_RESULT has the value 3 (allowed packet) in all cases and the signal PACKET_TYPE has the value 1 (TCP) in two cases and 2 (UDP) in the others. The generated statistics are also consistent with the test data, with four allowed packets, two tagged TCP and two tagged UDP.

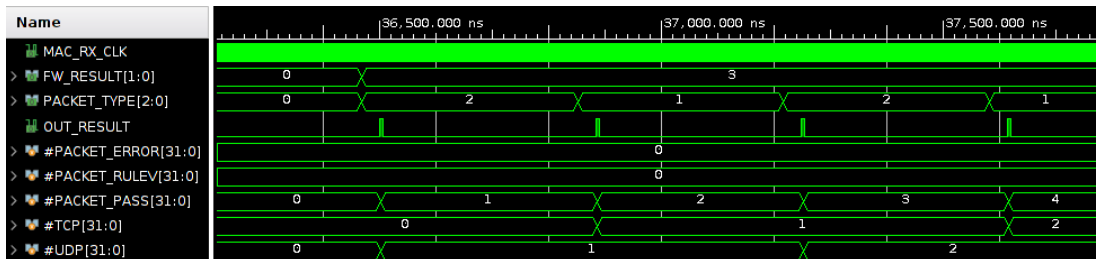
**Figure 5.** Signals waveform in the case of Ethernet frame rules check and statistics calculation.

## 3. Results

The proposed packet sniffer has been tested using the designed packet generator by feeding the system with a stream of Ethernet frames with controlled inter-frame delay. Tests were initially carried out to validate the packet sniffer in the case of 10 Gbit/s data rate. Then, the packet sniffer hardware was updated to work at the higher data rate of 100 Gbit/s and preliminary tests were carried out.

**3.1 Tests on 10 Gbit/s packet sniffer**

Different Ethernet frames (UDP protocol) of different lengths (50, 100, 200, 300, 400, 500, 750, 1000, and 1500 bytes) were tested with the packet generator working in loop mode. For each Ethernet frame, different values of the inter-frame delay were tested: 0, 1, 2, 3, 4, 5, 10, 50, 100, 500, 1000, 5000, 10000, and 50000 clock cycles, where each clock cycle has a period $T_{CLK}$ = 6.4 ns.

The Ethernet frames data rate (DR) can be expressed (in bit/s) as:

$$DR = \frac{64 \times N_{FW}}{(N_{FW} + N_{MAC} + N_{DELAY}) \times T_{CLK}} = 10^{10} \times \frac{N_{FW}}{N_{FW} + N_{MAC} + N_{DELAY}}$$

where $N_{FW}$ is the number of 8 bytes words of the Ethernet frame, $N_{DELAY}$ is the number of clock cycles of the inter-frame delay defined by us and $N_{MAC}$ is the number of clock cycles of delay between frames introduced by the Ethernet Subsystem IP. The test results are presented in Fig. 6, where the data rate (in Gbit/s) is plotted vs $N_{DELAY}$ for Ethernet frames of different lengths.

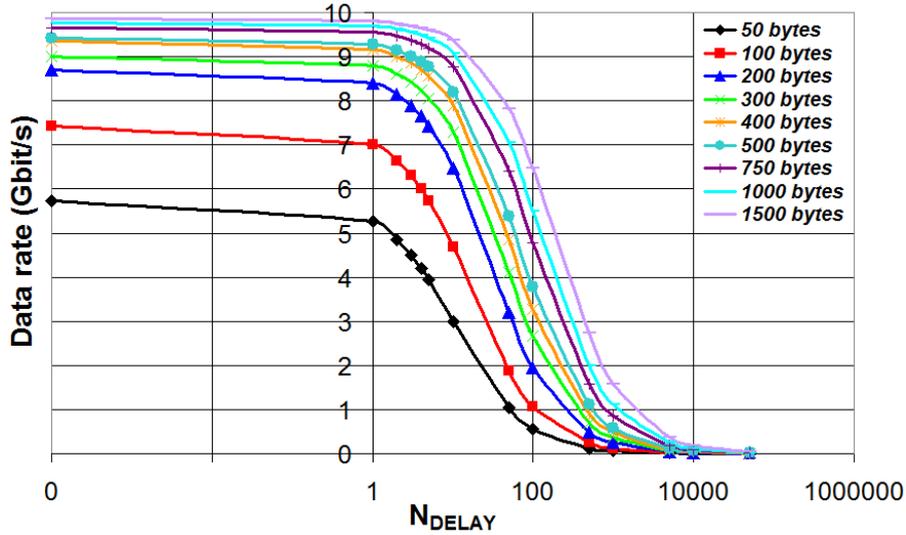

**Figure 6.** Data rate of Ethernet frames (in Gbit/s) plotted vs the number of clock cycles of introduced inter-frame delay for different UDP packets of different length.

As can be seen the data rate decreases with increasing the number of clock cycles of the inter-frame delay and the maximum data rate (obtained for $N_{DELAY}$ = 0) increases for higher frame length. This can be explained by the fact that the number of clock cycles lost due to Ethernet Subsystem IP delay $N_{MAC}$ impacts less on the data rate for long Ethernet frames and the theoretical maximum data rate of 10 Gbit/s is an asymptotic value for $N_{FW} \rightarrow +\infty$.

Then, the ability of the designed packet sniffer to discriminate between safe and potentially dangerous Ethernet frames has been tested. The packet generator was programmed to generate in loop mode sets of 1024 TCP packets with the source port variable in the range 0-1023. Six different packet transmissions, each of duration about 40 seconds, were carried out and, for each transmission, different firewall rules were loaded in the designed system to allow only TCP packets with the source port in the range 0-3, 0-15, 0-31, 0-63, 0-127 and 0-255. The results are presented in Fig. 7.

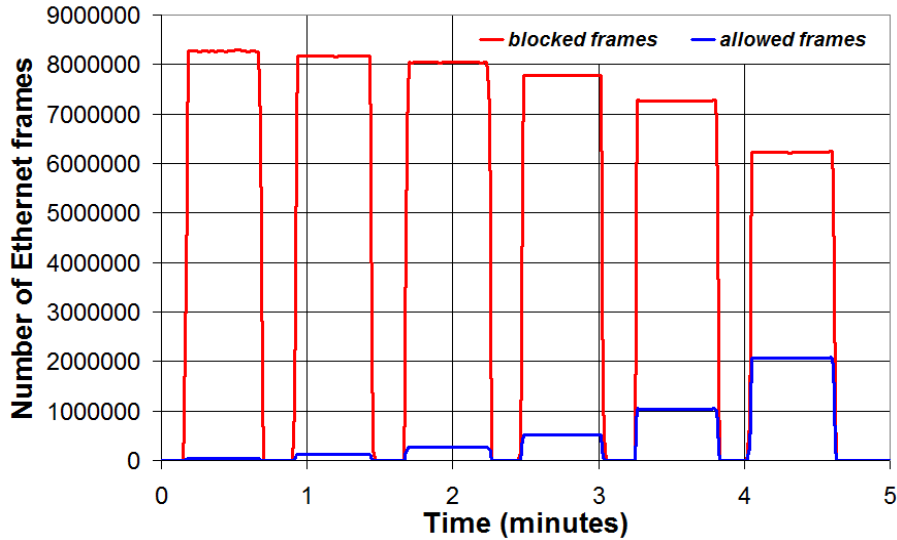

**Figure 7.** Number of allowed and blocked Ethernet frames in the case of TCP packets with different rules on the source port. The allowed source port range is 0-3, 0-15, 0-31, 0-63, 0-127 and 0-255 during the first, second, third, fourth, fifth and sixth packet transmission.

As expected, the number of allowed Ethernet frames is, respectively, 0.39%, 1.56%, 3.125%, 6.25%, 12.5%, and 25% of the total Ethernet frames generated. The choice of allowing Ethernet frames based on selected ranges of the source port represents only a case study to test the packet sniffer. The rules applied in a real world scenario are generally more complex and highly dependent on the application.

Overall, the tests on the designed packet sniffer have shown how the system can reliably monitor Ethernet traffic with data rates up to 10 Gbit/s.

### 3.2 Hardware upgrade to support 100 Gbit/s data rate

After validation of the 10 Gbit/s packet sniffer, the hardware was upgraded to allow the system to support the higher data throughput of 100 Gbit/s. In this regard, since the KC705 development board can support a maximum data throughput of 10 Gbit/s, a higher-performance FPGA device was used. In particular, the 100 Gbit/s packet sniffer was designed on the VCU1525 FPGA development board by Xilinx that integrates a VU9P Virtex UltraScale+ FPGA (featuring 2586 system logic cells, 6840 DSP slices, 345.9 Mbit of memory and 676 IOs), 64GB onboard DDR4 DIMM memory, two QSFP28 100G interfaces, PCI Express via Edge Connector and USB port for FPGA configuration via JTAG and UART communication [14].

As in the case of the 10 Gbit/s packet sniffer, a packet generator was designed to test the functionality of the packet sniffer. Since the VCU1525 development board features two QSFP28 100G interfaces, the packet generator was implemented on the same FPGA board as the packet sniffer. The physical layer of the 100Gbit/s Ethernet protocol was implemented using the UltraScale+ 100G Ethernet Subsystem IP provided by Xilinx (mode CAUI-4, 4 lanes at 25.78 Gbit/s). On the programmable logic side, Ethernet data are sampled with a clock of frequency 322.26 MHz and data width of 64 bytes (512 bits). The waveform signals are shown in Fig. 8 in the case of a received Ethernet frame of protocol TCP and length 88 bytes.

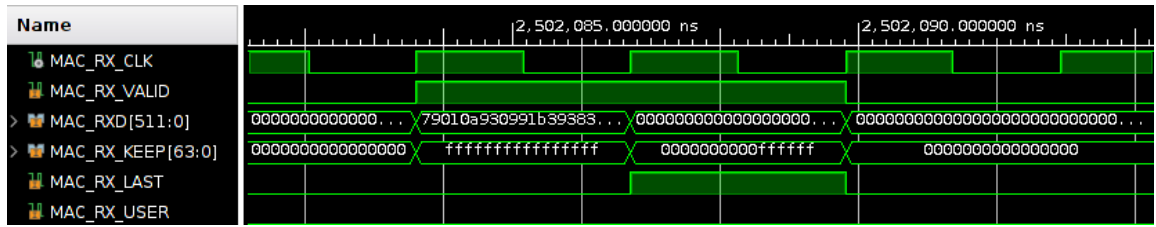

**Figure 8.** Waveform signals from a Vivado simulation in the case of a received Ethernet frame of protocol TCP and length 88 bytes.

As can be seen, the signal MAC_RX_VALID is active when valid data (MAC_RXD) are sampled and the signal MAC_RX_LAST is active during the last data transfer. The signal MAC_RX_KEEP (64 bit) indicates how many bytes are valid during the last data transfer. In the case of Fig. 8, for example, Ethernet frame data reading is carried out in two clock cycles: in the first cycle 64 bytes are received and in the second cycle 24 bytes are received, for a total of 88 bytes.

In Fig. 9 the waveform signals during the phase of Ethernet frame analysis are shown in the case of the same Ethernet frame of protocol TCP and length 88 bytes. As can be seen, the protocol of level 3 (LEV3_PROT) is 0x0800 (IP) and the protocol of level 4 (LEV4_PROT) is 6 (TCP).

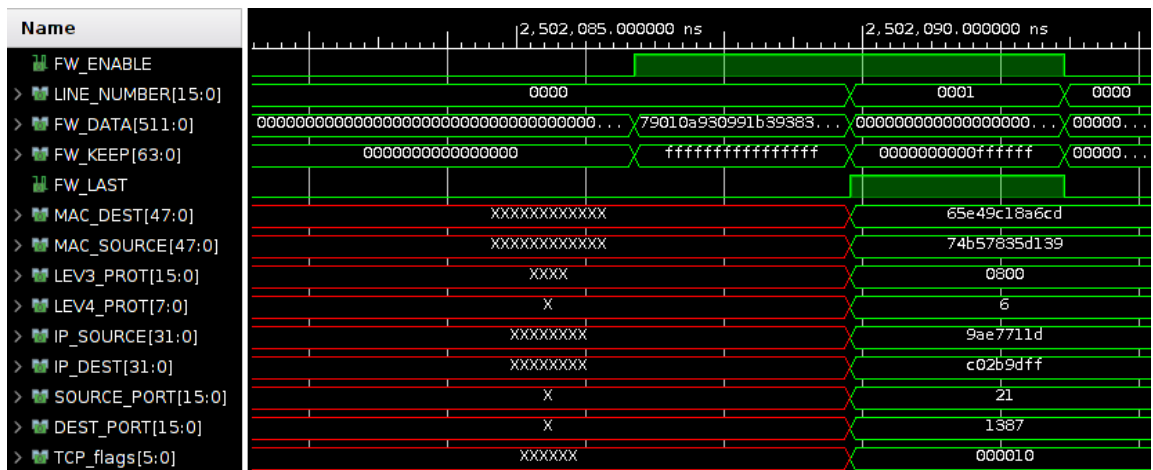

**Figure 9.** Waveform signals from a Vivado simulation in the case of the analysis of an Ethernet frame of protocol TCP and length 88 bytes.

As in the case of the 10 Gbit/s packet sniffer, also in the case of the 100 Gbit/s packet sniffer, tests were carried out using the packet generator with UDP packets of different length (64, 100, 200, 300, 400, 500, 750, 1000 and 1500 bytes) and different values of the inter-frame delay (0, 1, 2, 3, 4, 5, 10, 50, 100, 500, 1000, 5000, 10000 and 50000 clock cycles). Similarly with the case of the 10 Gbit/s packet sniffer, the data rate decreases with increasing the number of clock cycles of the inter frame delay and the maximum data rate (obtained for $N_{DELAY} = 0$) increases for higher frame length. The results are presented in Fig. 10.

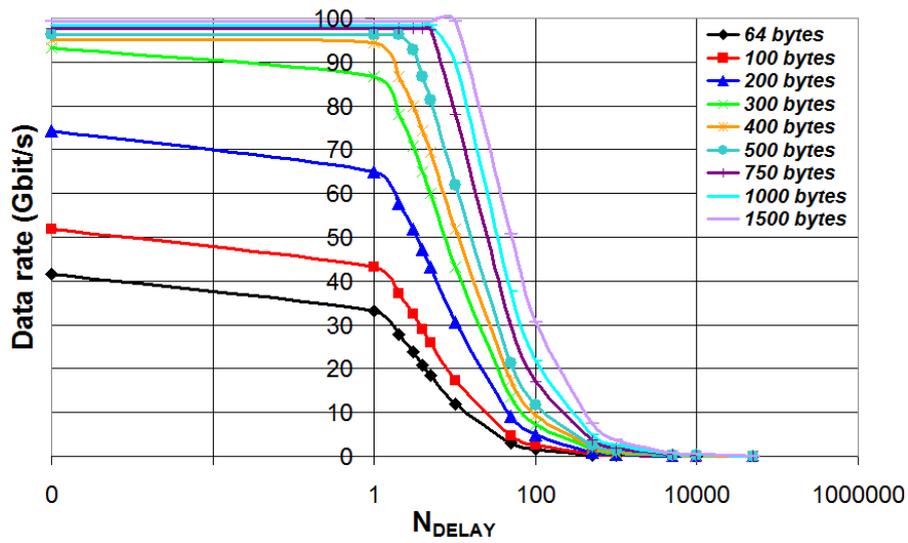

**Figure 10.** The data rate of Ethernet frames (in Gbit/s) plotted vs the number of clock cycles of introduced inter-frame delay for different UDP packets of different lengths in the case of the 100 Gbit/s packet sniffer.

The ability of the 100 Gbit/s packet sniffer to allow/block Ethernet frames based on a set of rules has been successfully tested only with single packets since, at the moment, the 100 Gbit/s packet generator has not the ability to generate streams of packets with iterations on source and destination ports.

## 4. Conclusions

A packet sniffer has been designed using FPGA to protect data transmitted using the Ethernet protocol. It is particularly suited to enhance data security in universities and research centers, where point-to-point network connections are dominant and large amount of sensitive data are shared among different hosts. The system operates on a stateless approach: when an Ethernet frame is received, important fields such as IP addresses, network and transport protocols, and source and destination ports are decoded and checked against a set of rules to detect potential threats. The system has been tested and validated for a data throughput of 10 Gbit/s with preliminary results showing that the data throughput can be reliably extended to 100 Gbit/s. In the end, the proposed device can significantly support the security of data transmission among research centers and decrease the chance of cyberattacks and data breaches.

## Funding

The Italian Ministry of University and Research, Grant/Award Number: J45F21002000001; "Alma Idea 2022" Linea di Intervento A (D.M. 737/2021); the Italian Ministry of Industry Incentives (MISE); and the Ministry of University and Research (MUR). In addition, this work was partially supported by project SERICS (PE00000014) under the MUR National Recovery and Resilience Plan funded by the European Union—NextGenerationEU.

## Acknowledgments

The authors would like to thank the National Institute for Nuclear Physics (INFN, Bologna division) and the National Center for Frame Analysis (CNAF, Bologna division) for their support in the development and testing of the presented packet sniffer.